\documentclass{article}
\usepackage{ltexpprt}
\usepackage{verbatim}
\usepackage{amssymb}
\usepackage{epsfig}
\usepackage{longtable}
\usepackage{tabularx}
\usepackage{amsfonts}
\def\C{\mathbb{C}}

\def\RR{\hbox{\vrule height 8pt width .7pt depth 0pt\hskip -.6pt\rm R}}
\def\C1{\hbox{{\rm C}\hskip -5pt\vrule height 7pt width .7pt depth -1pt\hskip 5pt}}
\renewcommand{\RR}{{\mathbb{R}}} 
 \newcommand{\Rmn}[2]{\RR^{#1 \times #2}}
\newcommand\exns{\begin{eqnarray}}
\newcommand\exne{\end{eqnarray}}
\newcommand\exs{\begin{eqnarray*}}
\newcommand\exe{\end{eqnarray*}}
\usepackage{pslatex}
\usepackage{verbatim}
\usepackage{amssymb}
\usepackage{graphicx}
\usepackage[bottom]{footmisc}

\usepackage{fancyhdr}
\begin{document}
\setlength\textwidth{5.125in}

\fontsize{12}{14}\selectfont
\lhead{\includegraphics[scale=0.3]{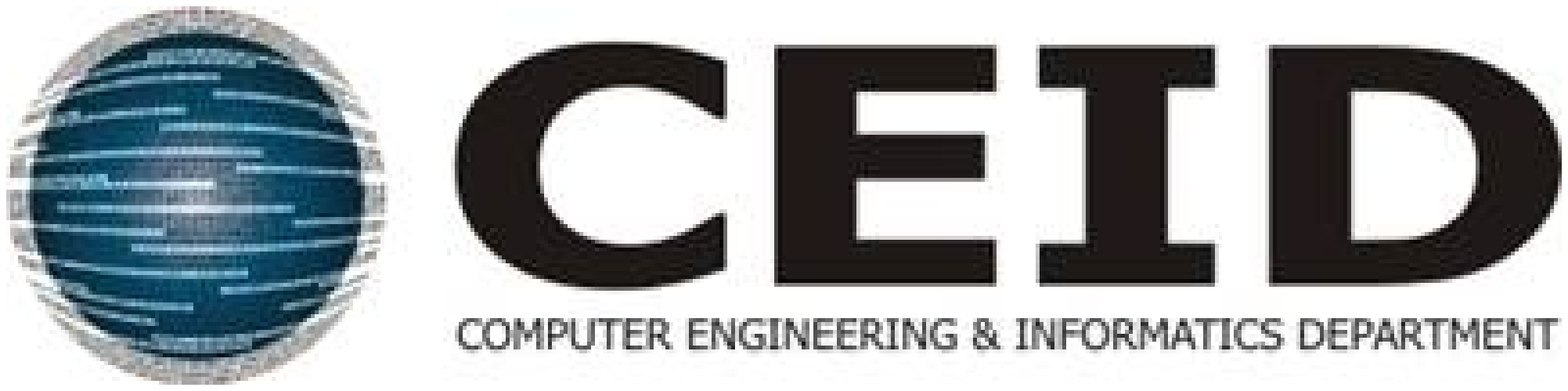}}
\rhead{\textsc{University of Patras\
}\includegraphics[scale=0.065]{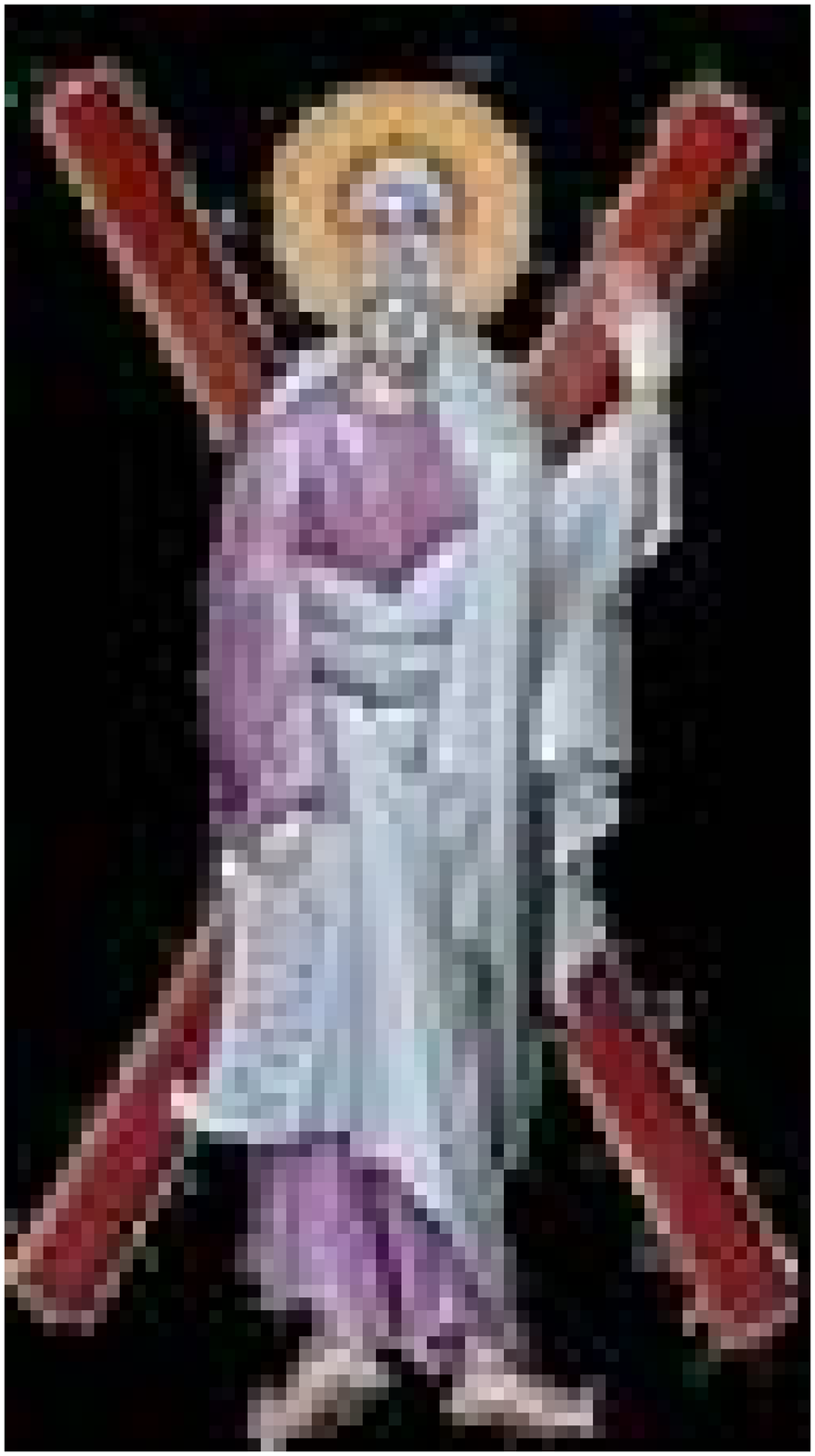}}
\lfoot{\tiny \textsc{Computer Engineering \& Informatics
Department, University of Patras, GR-26500, Patras, Greece}}
\rfoot{\tiny www.ceid.upatras.gr} \thispagestyle{fancy}
\renewcommand{\footrulewidth}{1.2pt}
\renewcommand{\headrulewidth}{1.2pt}
\newdimen\mytextwidth
\newdimen\myoddsidemargin
\newdimen\myevensidemargin
\mytextwidth=\textwidth \myoddsidemargin=\oddsidemargin
\myevensidemargin=\evensidemargin \setlength{\oddsidemargin}{0pt}
\setlength{\evensidemargin}{0pt}
\setlength{\textwidth}{1.15\textwidth} \thispagestyle{fancy}
\cfoot{}
\mbox{ } \\
\bigskip
\vskip 0.5in
\begin{center}
\begin{minipage}[c]{0.9\textwidth}
\textbf{\LARGE Exploring term-document matrices from matrix models
in text mining}
\end{minipage}
\end{center}

\bigskip
\begin{center} \textit{I. Antonellis} \textsl{and} \textit{E. Gallopoulos}
\end{center}

\bigskip
\begin{center}
February 2006
\end{center}

\vfill
\begingroup\small \flushleft
\begin{minipage}{\textwidth}
\begin{description}
\item[\bf Technical Report] \textsf{HPCLAB-SCG 03/02-06}
\item[\textsc{Laboratory:}] High Performance Information Systems Laboratory 
\item[\textsc{Grants:}] University of Patras ``Karatheodori''
Grant; Zosima Foundation Scholarship under grant
1039944/891/B0011/21-04-2003 (joint decision of Ministry of
Education and Ministry of Economics). \item[\textsc{Repository}]
\textsf{http://scgroup.hpclab.ceid.upatras.gr}
\item[\textsc{Ref:}] To appear in \textit{Proc. SIAM Text Mining
Workshop, SIAM Conf. Data Mining,} 2006. Supersedes
\textsf{HPCLAB-SCG 01/01-06}
\end{description}
\end{minipage}
\endgroup

\newpage
\mbox{ } \setlength{\textwidth}{\mytextwidth}
\setlength{\oddsidemargin}{\myoddsidemargin}
\setlength{\evensidemargin}{\myevensidemargin}
\fontsize{10}{12}\selectfont \setlength{\textwidth}{\mytextwidth}
\setlength{\oddsidemargin}{\myoddsidemargin}
\setlength{\evensidemargin}{\myevensidemargin}

\setcounter{page}{0} \twocolumn
\newpage
\textwidth 41pc

%


\title{Exploring term-document matrices from matrix models in text mining\thanks{Work conducted in the context of and supported
 in part by a University of Patras KARATHEODORI grant.}}
\author{ Ioannis Antonellis
\thanks{Computer Engineering and
Informatics Department, University of Patras, Greece. Supported in
part by a Zosima Foundation Scholarship under grant
1039944/891/B0011/21-04-2003 (joint decision of Ministry of
Education and Ministry of Economics). Email:
\texttt{antonell@ceid.upatras.gr}} \\
\and Efstratios Gallopoulos \thanks{Computer Engineering and
Informatics Department, University of Patras, Greece. Email:
\texttt{stratis@ceid.upatras.gr}}}

\date{}

\maketitle


\begin{abstract} \small

We explore  a matrix-space model, that is a natural extension to
the vector space model for Information Retrieval. Each document
can be represented by a matrix that is based on document extracts
(e.g. sentences, paragraphs, sections). We focus on the
performance of this model for the specific case in which documents
are originally represented as term-by-sentence matrices. We use
the singular value decomposition to approximate the
term-by-sentence matrices and assemble these results to form the
pseudo-``term-document'' matrix that forms the basis of a text
mining method alternative to traditional VSM and LSI. We
investigate the singular values of this matrix and provide
experimental evidence suggesting that the method can be
particularly effective in terms of accuracy for text collections
with multi-topic  documents, such as web pages with news.
\end{abstract}

\section{Introduction \label{sec:1}}


The vector space model (VSM), introduced by Salton
\cite{SaltonWongYang.75}, is one of the oldest and most
extensively studied models for text mining. This is so because it
permits using theories and tools from the area of linear algebra
along with a number of heuristics. A collection of $n$ documents
is represented by a term-by-document matrix (tdm) of $n$ columns
and $m$ rows, where $m$ is the number of terms used to index the
collection. Each element $a_{ij}$ of the matrix is a suitable\
measure of the importance of term $i$ with respect to the document
and the entire collection. Although numerous alternative weighting
schemes have been proposed and extensively studied, there are some
well-documented weaknesses that have motivated the development of
new methods building on VSM. The best known is  Latent Semantic
Indexing (LSI) \cite{LSIproto}, where the column space of the tdm
is approximated by a space of much smaller dimension that is
obtained from the leading singular vectors of the matrix. The
model is frequently found to be very effective even though the
analysis of its success in not as straightforward
\cite{PapLSI.00}. The computational kernel in LSI is the singular
value decomposition (SVD) applied on the tdm. This provides the
mechanism for projecting  data onto a lower, $k$-dimensional space
spanned by the $k$ leading left singular vectors; cf. the
exposition in \cite{Berry.LSI.Book, Berry.Browne.99,
EldenActa.06}.
    In addition to performing dimensionality reduction, LSI  captures hidden semantic structure in the data and resolves
problems caused in VSM by synonymy and polysemy. A well-known
difficulty with LSI is the cost of the SVD for the large, sparse
tdm's appearing in practice. This complicates not only the
original approximation but also the updating of the tdm whenever
new documents are to be added or removed from the original
document collection.
These are obstacles to the application of LSI on very large tdm's,
so several efforts in the area are directed towards alleviating
this cost. These range from  techniques for lowering the cost of
the (partial) SVD (e.g. exploiting sparse matrix technology and
fast iterative methods, cf.
\cite{svd-berry,BlomRuhe.05,simon00lowrank}), to the application
of randomized techniques (\cite{achlioptas,
drineas.frieze.kannan.vempala}) specifically targeting very large
tdm's. One approach that appears to be promising is to approximate
the tdm by operating on groups of documents that either arise
naturally (e.g. because the documents reside at distant locations)
or as a result of clustering
\cite{CastelliThomasian.03, kargupta01distributed,
 clsi-SIAM.DM05}. It was shown in
\cite{clsi-SIAM.DM05}, for example, that by clustering and then
using few top left singular vectors of the tdm corresponding to
each cluster could lead to economical and effective approximation
of the tdm.

In this ``work in progress'', we explore a family of text mining
models arising as a natural extension of VSM and present cases
where they   appear to be able to capture more information about
text documents and their structure. Our starting point is that the
tdm utilized in VSM and LSI has no ``memory'' how it was
constructed; in particular, any of the tdm columns can be
decomposed in an unlimited number of ways as linear combination of
other vectors. We can, however, express each document vector as
the sum of vectors resulting from the document terms appearing at
a selected level of the
 document's hierarchical structure (e.g. sections, paragraphs, sentences etc.) We can thus consider each
 of the document vectors to be the product of a ``term-by-extract unit'' matrix with a vector of all 1's.
 Based on this, it is a natural next step to consider approximating each term-document vector.
 We would be loosely referring to the general idea as
 \textit{Matrix Space Model}
(MSM). MSM  permits us to capture suitable decompositions of
document vectors based on the document's hierarchical structure
(into sections, paragraphs, sentences etc.) and store them into a
matrix. To explore the model's properties, we study a specific
instance  based upon document decomposition into sentences.
Sentence based decompositions have  already been  applied in text
classification and summarization \cite{antonellis.bouras.06,
Castellanos.04, SCOL03,WuGunopulos02},
 therefore the analysis we provide
is also of independent interest. We also note an elegant recent
proposal for a matrix-based IR framework close, but not the same,
as ours in \cite{Roelleke-etal2004} as well as another
phrase-based framework \cite{HammoudaKamel.04} for clustering of
semi-structured Web documents. We discuss these approaches later
in this paper. As will become apparent in the sequel, one common
useful feature of MSM-type models is that they can  readily lead
to the tdm of the original VSM.

Based on the representation of the tdm as a matrix whose columns
are obtained by multiplying a ``term-by-extract unit'' matrix with
a vector of all 1's, we  approximate each column based on this
decomposition. It is worth noting that our proposed approach has
an analogue in the numerical solution of partial differential
equations, namely domain decomposition techniques based on
substructuring \cite{BjWi86}. These are powerful tools that also
lend themselves to parallel processing.

 The rest of the paper is structured as follows.
In Section \ref{sec:2} we describe the matrix space models and
show their relation with
 classic VSM and its variants. We also provide a formal study of the text analysis using document decomposition into its sentences and introduce
formal definitions for the term-by-sentence and other matrices
useful for MSM. Based on these, we describe a general IR method
based on this approach and specify its use for the case of
sentence-based analysis.
In Section \ref{sec:5} we analyze the method and its relevant
costs, and derive spectral information for the
matrix underlying the IR strategy. 
Section \ref{sec:7} presents our experimental analysis. Finally in
Section \ref{sec:8} we give our conclusions and future directions.

Throughout the paper, we  use pseudo-MATLAB notation. We would be
referring to the $j$-th column of any matrix $A$ as
 $a_{j}$, so that  $a_{j} = A e_{j}$, where $e_{j}$ denotes the $j$-th column of an (appropriately sized)
indentity matrix $I$. We will also represent $A$ as
\begin{eqnarray}
    A = \left[a_{1},  \dots, \dots\ a_{m} \right].
    \nonumber
\end{eqnarray}
 and use ${\rm best}_{k}\left(A\right)$ to denote its best rank-$k$ approximation. We will use $e$ to refer to the vector of
all 1's, whose size is assumed to be appropriate for the
computation to be valid. Given scalars (square submatrices)
$\phi_j$, we use ${\rm diag}[\phi_1, ..., \phi_j]$ to denote the
corresponding diagonal (block diagonal) matrix. When the need
arises (e.g. two ``$e$''-vectors of different dimension in the
same formula) a superscript will be used to show the difference,
e.g. $e^{(k)}$.

\section{Matrix space models for IR \label{sec:2}}

In the VSM, each  document is represented using an $m$-dimensional
vector. Each of the $m$ dimensions refers to an indexing term and
each coordinate of the vector is computed using some combination
of a local and/or global weighting scheme. Weighting schemes can
be seen as heuristics that help eliminate problems arising from
the non-orthogonality of the different indexing terms and have
been proven to be efficient for improving ``precision'' and
``recall''.

We next observe that for a vector representation of a document,
there are unlimited decompositions into a (given) number of
components. Using the vector space model, such components could be
seen as different ``concepts'', which combined, generate the
concept of the given document. In fact, some reasonable
decompositions would create the components using consecutive
document's extracts. As VSM stores only the final vector of each
document, it is obvious that it doesn't exploit such kind of extra
information. The goal of MSM's is to utilize meaningful document
decompositions that are based upon its structure; cf.
\cite{HammoudaKamel.04,Roelleke-etal2004}. As document structure
often builds a hierarchy into sections, paragraphs and sentences,
we can decompose each document into the vector space
representation of non-overlapping and sequential extracts that
correspond to them and store such decompositions into a
\textit{term-by-extract} matrix (tem) (also called
``term-location'' matrix in \cite{Roelleke-etal2004}.) The choice
of the hierarchy's level that the decomposition will rely on, can
result in document representation using \textit{term-by-section}
(tsm), \textit{term-by-paragraph} (tpm) or
\textit{term-by-sentence} (tsm) matrices. The common thread is
that MSMs use matrices to store vector space representations of
document's extracts. The $j$-th column of such a matrix refers to
the vector space representation (based only on term frequency) of
the $j$-th extract of the document.
These are features that our paper shares with
 \cite{HammoudaKamel.04,Roelleke-etal2004}. On the other hand,
\cite{HammoudaKamel.04} addresses primarily the issue of effective
indexing - via subgraphs and document index graphs - for
sentence-based analyses, while \cite{Roelleke-etal2004} is
concerned with the formal framework surrounding term-location and
term-document matrices. None of these papers, however, considers
the idea proposed herein, namely the replacement of the original
document vectors with approximants and the effect of such
replacements on retrieval performance.

\begin{figure}[htbp]
\begin{center}
\includegraphics[scale=.40,angle=0]{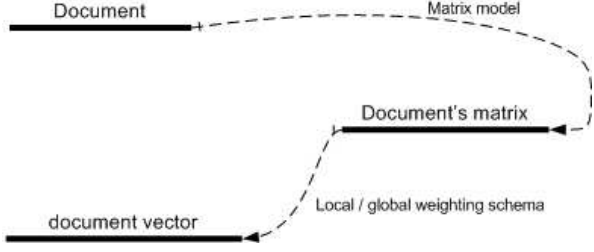}
\caption{ {\small Transition between matrix space and vector
space.} } \label{transition_without_svd}
\end{center}
\end{figure}

As Figure \ref{transition_without_svd} illustrates, MSM can be
used as a transitional phase before producing the document vector.
Given a tem $H \in \mathbb{R}^{m \times n} $ of a document $D$, we
can construct its vector space representation that is based on an
arbitrary combination of local and/or global weighting scheme. As
elements $h_{ij}$ of $H$ are based only on term frequency, the
vector space representation of any document vector $a\in
\mathbb{R}^{m}$ in the tdm can be written as $a = G H p$, where $G
\in \mathbb{R}^{m \times m}$ and $p \in \mathbb{R}^{n}$. Matrix G
is a diagonal matrix with nonzero diagonal elements accounting for
the global weighting scheme. Column vector $p$ corresponds to the
local weighting scheme applied on document $D$. For example, $G =
I$ corresponds to the application of no global weighting scheme,
while $p = e$ corresponds to ``term frequency'' local weighting.
Hereafter, we will assume that transition from matrix space to
vector space is done by applying $G = I$ and $p = e$ to the tem.
However, our results can be generalized to include more complex
weighting schemes.

In the following section, we study MSM based upon document
representation using tsm's. For simplicity, we define a
``sentence'' to be text delimited between two consecutive periods
(``.''). We do not address here the interesting issues involved in
sentence identification (e.g. see \cite{Castellanos.04,
palmer.sentence94}.)

\subsection{Text analysis based on sentences. \label{sec:3}}
Let $A$ denote a tdm of rank $r$ and let its SVD
  be
\begin{equation}
    A = U \Sigma V^\top = \sum_{i=1}^{r} \sigma_{i} u_{i} v_{i}^{\top},
\label{*eq*:svdofa}
\end{equation}
where the rightmost expression is the dyadic decomposition and, as
usual, the singular values are arranged in non-increasing order.
We also write ${\rm best}_k(A)$ for the best rank-$k$
approximation of $A$ (we assume here that $\sigma_k
> \sigma_{k+1}$):
\begin{equation}
   { \rm best}_k(A) = \sum_{i=1}^{k} \sigma_{i} u_{i} v_{i}^{\top}
\label{*eq*:svdofak}
\end{equation}
Note also that
\begin{equation}
\begin{array}{rcl}
    a_{j} &=& \left(\sum_{i=1}^{r} \sigma_{i} u_{i} v_{i}^{\top}
    \right)_{j}
    = \sum_{i=1}^{r} \sigma_{i} u_{i} v_{ij}
\label{*eq*:svdofarevised}
\end{array}
\end{equation}
and similarly for  the $j$-th column of  ${\rm best}_k(A)$:
\begin{equation}
    {\rm best}_k(A)e_j = \sum_{i=1}^{k} \sigma_{i} u_{i} v_{ij}
\label{*eq*:svdofaêrevised2}
\end{equation}
Having assumed that matrix $A$ is a tdm, the $j$-th column of $A$
will correspond to the vector space representation of the $j$-th
document of the collection. So,
we can write
\begin{equation}
    a_{j} = \sum_{k=1}^{m} s_k,
\label{*eq*:aidecompotition}
\end{equation}
where $s_k$,  $k=1 \dots m$, is the vector space representation of
the $k$-th sentence of document $j$ of the collection and $m$ the
total number of sentences of the $j$-th document. We can now
construct the tsm of document $j$ of our collection according to
the following definition:
\begin{definition}
Let document $D$ contain $m$ sentences and $d$ be its vector space
representation. The term-by-sentence matrix of document $D$ is the
matrix
\begin{equation}
    S_D = \left[ s_{1}, s_{2}, \dots, s_{m} \right],
\label{*eq*:sdefinition}
\end{equation}
where $s_k$ refers to the vector space representation of the
$k$-th sentence of $D$.
\end{definition}
Using the above notation, Equation \ref{*eq*:aidecompotition} can
be written as
\begin{equation}
    a_{j} = A e_j = S_D e^{(m)}.
\label{*eq*:aidecompotitionrevised}
\end{equation}
We also introduce the  notion of the ``term-by-sentence matrix for
a matrix collection''. For example, if we have two documents $D_1,
D_2$, their tsm's are $S_{D_1}$ and $S_{D_2}$, and the usual tdm
from the VSM is the matrix of two columns $A = [a_1,a_2]$, then
the tsm for the collection
 is the matrix
$S_C = [\widehat{S}_{D_1}, \widehat{S}_{D_2}]$, where
$\widehat{S}_{D_j}$ is an embedding of the original $S_{D_j}$ into
a matrix with as many rows (terms) as $A$. In other words, we
augment each one of the tsm's $S_{D_j}$ with zero rows
corresponding to those terms in the collection's vocabulary but
not present in $D_j$. In general, we have the following:
\begin{definition}
Let $C$ be a collection of $k$ documents $D_1,\ D_2,\ \dots\ D_k$,
where the $i$-th document consists of $m_i$ sentences. The
term-by-sentence matrix of the collection $C$ is the matrix $S_C$:
\begin{equation}
\begin{array}{rcl}
    S_C &=& \left[ \widehat{S}_{D_1},  \widehat{S}_{D_2}, \dots\, \widehat{S}_{D_k}\right]\\
\end{array}
\label{*eq*:sdefinition2}
\end{equation}
where $\widehat{S}_{D_j}$ is an embedding of the original
$S_{D_j}$ into a zero matrix with as many rows as the tdm of the
VSM representation for $C$.
\end{definition}



%
%
The MSM provides a more general framework  for IR
\cite{HammoudaKamel.04,Roelleke-etal2004}. Our objective, here,
was to investigate the  performance  of such a scheme and evaluate
it relative to LSI and VSM. As we show, in specific cases the
method can  achieve results similar to LSI with respect to
accuracy measures such as precision and recall, while keeping the
computational costs to the levels of simple VSM.

The rationale of our method is that by projecting sentence vectors
of tsm's onto the subspace spanned by the singular vectors
corresponding to the $k$ largest singular values for some small
value of $k$, permits us to eliminate polysemy and synonymy
phenomena within the document. This is accomplished by the rank
reduction of tsm's that SVD produces. It is obvious that, as these
phenomena are eliminated locally in every document
it will be difficult to improve on LSI.


However, when the method is applied to collections with documents
whose context is not semantically specific but multi-topic (e.g.
documents from web pages with news) or to collections with large
percentage of different terms per document, we provide
experimental evidence that its performance surpasses classic
vector space and comes close to LSI's performance. The objective
is, when a document that refers to $k$ semantically well-separated
topics is given as input, for the projection to identify the
principal directions of these topics. Then, by using the projected
sentence vectors (instead of the original ones) we can transition
to the vector space (from the MSM that uses tsm) by constructing
approximations to the vectors of the VSM tdm according to some
weighting scheme.
We would thus be referring to this  tdm with approximated columns
as ``pseudo-tdm''. In particular, the $j$-th column of matrix $A$
is not computed according to Equation
\ref{*eq*:aidecompotitionrevised} but as follows: Let $D_j$ the
$j$-th document, $S$ the corresponding tsm, $r$ its rank and $t$
the number of columns (sentences) of in $D_j$. Then
\begin{equation}
 \ \    S = \sum_{i=1}^{r} \sigma_{i} u_{i} v_{i}^{\top}\ \textrm{and} \ {\rm best}_{k'}(S) = \sum_{i=1}^{k'} \sigma_{i} u_{i} v_{i}^{\top},
\label{*eq*:sentencesvd}
\end{equation}
where $k' \leq r$, $(\sigma_i, u_i, v_i)$ are singular triplets of
$S$ with the $\sigma_i$'s arranged in decreasing order, and
 $\sigma_{k'} > \sigma_{k'+1}$.
%
%
Then column $a_j$ of the pseudo-tdm $A$ can be constructed as
\begin{equation}
   a_j = {\rm best}_{k'} (S) e^{(t)}.
\label{*eq*:sentencesvdk}
\end{equation}

The steps for the sentence-oriented algorithm we are experimenting
with are shown in Table \ref{tb:Algo}.
\begin{table}
\begin{center}
\begin{tabular}{l}\hline
{\bf Algorithm: Construct pseudo-tdm based on tsm} \\
{\bf Input}: Document collection $\{D_1, ..., D_m\}$ \\
{\bf
Output}: Pseudo-tdm $A$ \\
I. For each document $D_j$: \\
\ \ \ 1. Prepare tsm $S_{D_j}$ \\
\ \ \ 2. Select $k' \leq {\rm rank}(S)$ \\
\ \ \ 3. $a_j = {\rm best}_{k'} (S) e^{(t)}$ \\
II. Assemble pseudo-tdm $A := [a_1, ..., a_m]$ \\
\hline
\end{tabular}
\caption{Construction of pseudo-tdm from document
collection.}\label{tb:Algo}
\end{center}
\end{table}
Query vectors $q$ are therefore compared to the columns of the
pseudo-tdm. Cosine similarity can be computed using the following
formula:
\begin{equation}
\begin{array}{rcl}
    \cos\ \theta_{j} &=& \frac{\left( Ae_j\right)^\top q}{\|Ae_j\|_2\|q\|_2}\\
    &=& \frac{\left(\sum_{t=1}^m \sum_{i=1}^{k'} \sigma_{i} u_{i}^\top v_{i,t}\right)^\top q}{\|Ae_j\|_2\|q\|_2}
\label{*eq*:cosines}
\end{array}
\end{equation}

\begin{figure}[htbp]
\begin{center}
\includegraphics[scale=.35,angle=0]{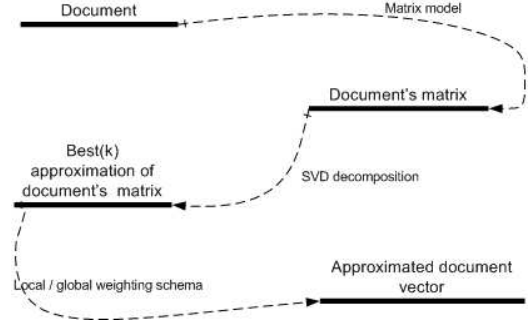}
\caption{ {\small Using MSM as a tool for developing a new IR
method.} } \label{transition_with_svd}
\end{center}
\end{figure}

Figure \ref{transition_with_svd} depicts schematically how MSM can
be used to develop a new IR method. Transitional representation of
a document in the matrix space permits the application of matrix
transformations (in our case,  approximation via SVD) before
producing the vector that will represent the document in the tdm.
Each approximated document vector is ``assembled'' from structures
that are local to the document (substructures).

It is worth noting that in the algorithm presented in Table
\ref{tb:Algo}, each $k'$ is selected to be $k' \leq r$, where $r$
is the rank of the particular tsm. Therefore, if we choose $k' =
r$,
 the resulting vector becomes identical to the one obtained in the tdm of the classical VSM (before
any weighting); furthermore, there is no need to perform an SVD of
the tsm. Therefore, if this choice is made for every document,
 the whole pseudo-tdm reverts to the usual tdm, highlighting  that MSM  is a generalization of VSM.

\subsection{Computational costs.} As with LSI, the method also relies on the (partial) SVD. The difference with LSI
is that there are multiple SVD's, one per tsm for each document.
Note that the tsm's can differ widely  in size.
 Furthermore, the number of rows of each tsm will typically be much smaller
than the full tdm since the number of terms in each sentence is
expected to be much smaller than the total number of terms in the
collection. More importantly, even though the approximate
term-document vector resulting from this process for each document
might be far less sparse than the term-document vector
corresponding to the original tdm, many zeros will be introduced
at the embedding phase, to take into account terms that are
present in other documents but not this one.
It is also worth noting, though we leave it for future study, that
the
 SVD's are independent for each document and hence  can be processed in a distributed manner.
It can happen, of course, that the number of sentences in each
document can be large, even larger than the number of documents in
the entire collection. To address this, we can make use of the
flexibility of the proposed methodology, and adjust our analysis
at any level of the document's hierarchy that is convenient
(sentences, paragraphs, sections, ...). This is work in progress
and we plan to report on it in the future. Table
~\ref{datasetsizes} illustrates indicative sizes for the tsm's
resulting from the MEDLINE datasets. These tsm's appear to be
small enough compared to the tdm and therefore the application of
our method to them results on computational costs very close to
that of VSM. We finally note that another advantage of this
approach is that as the approximation is performed locally for
each document, the method does not entail significant costs when
performing document updates. In particular, the update of the
pseudo-tdm using a new document  will only cause a non-trivial
change on existing document vectors if terms that already existed
in these documents but were not accounted for till then, e.g.
because of low global frequency. In that case, we would need to
update, in some way, the SVD of the tsm of each affected document.
\begin{center}
 \begin{table}[tbh]
\begin{center}
 \begin{tabular}{|c||c|c|c|c|}
    \hline
MED &  terms   &  sentences & total & total \\
\#&  per doc. &  per doc. & terms & docs.\\
    \hline
1 & 45 (0.81\%) & 6 (0.58\%) & 5526& 1033\\
2& 90 (1.62\%) & 13 (0.62\%)  & 5526&  2066\\
3& 135 (2.44\%) & 19 (0.61\%)  & 5526& 3099\\
4& 180 (3.25\%) & 24 (0.58\%)  & 5526& 4132\\
5& 225 (4.07\%) & 28 (0.54\%)  & 5526& 5165\\
6& 270 (4.88\%) & 34 (0.55\%)  & 5526& 6198\\
7& 315 (5.70\%) & 40 (0.55\%)  & 5526& 7231\\
8& 360 (6.51\%) & 47 (0.57\%)  & 5526& 8264\\
9& 405 (7.32\%) & 52 (0.56\%)  & 5526& 9297\\
10& 450 (8.14\%) & 58 (0.56\%)  & 5526& 10330\\
    \hline
\end{tabular}
\end{center}
 \caption{Example sizes for tsm's (average number of terms and sentences per document for the datasets we used). }
 \label{datasetsizes}
 \end{table}
\end{center}

%

\section{Analysis  \label{sec:5}}
To shed further light into the nature of the matrix resulting from
tsm's, in this section, we study the behavior of the singular
values of a collection's pseudo-tdm that has been constructed
using this approach. We  show that pseudo-tdm's preserve the
so-called \textit{low-rank-plus shift} property of
tdm's(\cite{SimonZha.99}). Note that this property can be put to
practical use to enhance the effectiveness of update algorithms
(\cite{zha.zhang.low00}). Even though we do not study such update
schemes here, it is useful to know that they could be applied on
pseudo-tdm's.


\begin{definition} [Low rank plus shift structure]
A matrix $A \in \Rmn{m}{n}$ is said to have  ``low rank plus
shift'' structure if it satisfies:
\begin{equation}
    \frac{A^\top A}{m} \approx CWC^\top + \sigma^2 I
     \label{*eq*:definitionlowrankplusshift}
\end{equation}
where $C \in \Rmn{n}{k} $ and matrix $W \in \Rmn{k}{k}$ is a
symmetric positive definite matrix.
\end{definition}
When $A$ represents a collection's tdm, $C \in \Rmn{n}{k}$ is the
matrix whose columns represent latent concepts of the collection.
The use of the terminology ``low rank plus shift'' comes from the
fact that in IR applications, $k \ll \min\{m,n\}$. The singular
values of such matrices have the following distribution: The first
few  singular values are large but decrease rapidly and then the
curve becomes flat but not necessarily zero. In order to determine
if a matrix satisfies this property, we follow the analysis of Zha
and Zhang (\cite{zha.zhang.low00}) and investigate the following
matrix approximation problem: Given a rectangular matrix, what is
the closest matrix that has the low rank plus shift property. We
can then define a matrix set for a given $k > 0$,

\begin{eqnarray}
    J_k = \{ B \in \Rmn{m}{n} \  |\  \sigma_1(B)\geq \ldots \geq \sigma_{\min\{m,n\}}(B), \nonumber \\
     \sigma_{k+1}(B) = \cdots = \sigma_{\min\{m,n\}}(B)\}
\end{eqnarray}
Using this notation, the matrix approximation problem is reduced
to finding the distance between a general matrix $A$ and the set
$J_k$. The next theorem provides as such a solution.
\begin{theorem} [Zha and Zhang \cite{SimonZha.99}]
Let the SVD of $A$ be $A = U\Sigma V^\top$, $\Sigma = {\rm
diag}[\sigma_1,\ldots ,\sigma_{\min\{m,n\}}]$ and $U,\ V$
orthogonal. Then for $k \leq \min\{m,n\}$ we have:
\begin{eqnarray}
    \arg\ \min_{J \in J_k} \|A - J\|_p = U_k \Sigma_k V_k^\top +
    \tau_p U_ê^\bot(V_k^\bot)^\top, \nonumber
\end{eqnarray}
where $\Sigma_k ={\rm diag}(\sigma_1, \ldots, \sigma_k), U = [U_k,
U_k^\bot]$, $V = [V_k, V_k^\bot]$ and $p$ refers to either the
Frobenius ($p=F$) or the spectral ($p=2$) norm. Furthermore,

\begin{eqnarray}
    \tau_F =  \sum_{i=k+1}^{\min\{m,n\}}\frac{\sigma_i}{(\min\{m,n\} -
    k)}\nonumber
\end{eqnarray}
and
\begin{eqnarray}
    \tau_2 =  \frac{\sigma_{k+1} +\sigma_{\min\{m,n\}}}{2} \nonumber
\end{eqnarray}
\label{theorem:lowrank}
\end{theorem}
Using the above theorem, we can examine experimentally how close
is a given tdm to the set of matrices with the ``low rank plus
shift'' structure. For our experimental analysis, we used the
MEDLINE dataset that contains relatively small documents and one
topic per document. We also constructed artificial, additional
datasets based on MEDLINE, so as to test the performance of our
method when applied to multi-topic documents. The documents of
these datasets (MED\_1, MED\_2, ..., MED\_10) consist of joint
documents of original MEDLINE. Documents of MED\_i dataset,
contain $i$ MEDLINE documents; MED\_1 is identical to MEDLINE.

Table \ref{*table*:lowrank} shows the value of quantity $\frac{\|A
-{\rm best}_{100}(A)\|_F}{\|A\|_F}$ for different tdm's
(pseudo-tdm and the usual VSM tdm). According to Theorem
\ref{theorem:lowrank}, the smaller this value is for a given
matrix, the closer to low-rank-plus-shift structure the matrix is.
The notation we use for the naming of the datasets is of the form
NAME\_i\_j, where NAME is the dataset's name, $i$ is the number of
actual semantic topics per collection's document and $j$ is the
number of singular triplets of the term by sentences matrix that
were used for the construction of approximated document vectors.
\begin{table}[tbh]
 \begin{center}
 \begin{tabular}{|c|c|c|}
    \hline
Dataset &  VSM tdm &   pseudo-tdm \\
    \hline
MED\_1\_1   & $0.6738$ &  $0.6655$\\
MED\_2\_1   & $0.5973$ &  $0.5807$\\
MED\_3\_1   & $0.5232$ &  $0.5030$\\
MED\_4\_1   & $0.4601$ &  $0.4379$\\
MED\_5\_1   & $0.4015$ &  $0.3746$\\
MED\_6\_1   & $0.3462$ &  $0.3172$\\
MED\_7\_1   & $0.2852$ &  $0.2534$\\
MED\_8\_1   & $0.2282$ &  $0.1982$\\
MED\_9\_1   & $0.1644$ &  $0.1411$\\
MED\_10\_1  & $0.0829$ &  $0.0660$\\
    \hline
\end{tabular}
\end{center}
 \caption{Value of $\frac{\|A -{\rm best}_{100}(A)\|_F}{\|A\|_F}$ for a tdm
 constructed using VSM and our method (pseudo-tdm).
}
 \label{*table*:lowrank}
 \end{table}

\begin{figure}[htbp]
\begin{center}
\includegraphics[scale=.30,angle=0]{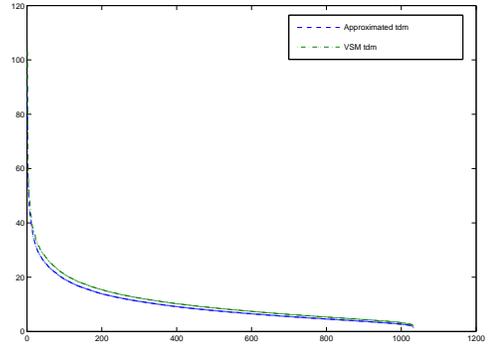}
\includegraphics[scale=.30,angle=0]{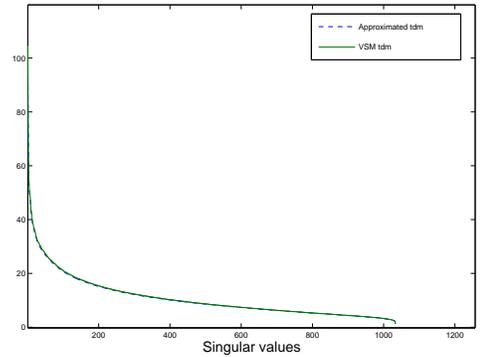}
\includegraphics[scale=.30,angle=0]{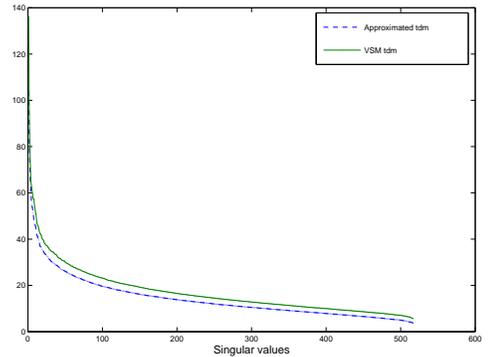}
\includegraphics[scale=.30,angle=0]{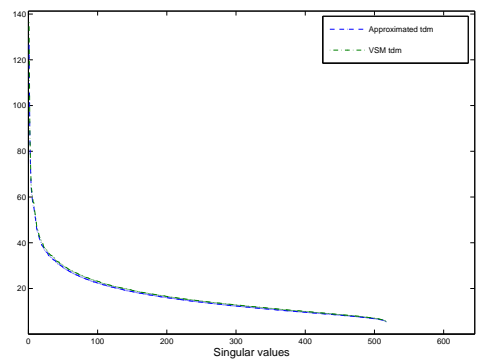}

\caption{ {\small Plots of singular values  for the classic and
the approximated tdm that arises when one singular value of term
by sentences matrix for MEDLINE-1 dataset (first), five singular
values for MEDLINE-1 dataset (second), one singular value for
MEDLINE-2 dataset (third) and five singular values for MEDLINE-2
dataset were used.} } \label{FS1}
\end{center}
\end{figure}

As depicted on Table ~\ref{*table*:lowrank}, the pseudo-tdm's
appears to be closer to the  ``low rank plus shift'' structure
than the  VSM tdm's. Furthermore, the distance becomes even
smaller for
 multi-topic collections. Figure \ref{FS1}
depicts singular value distributions for classic and approximated
tdm's of datasets ${\rm MED}\_1$ and ${\rm MED}\_2$ using $k'=1,
5$ singular triplets to approximate the tsm's. We note that the
singular values of pseudo-tdm's are bounded by the corresponding
singular values of VSM tdm's. We next prove that this indeed
holds.

\subsection{Some spectral properties of pseudo-tdm's.}

%
In the sequel, when we  compare the singular values of two
matrices with the same number of rows but different number of
columns we will count the singular values according to the number
of rows. We first state three simple results.
\begin{lemma}\label{lem:singular_orthonormal}
Let $A \in \Rmn{m}{n}$. Let V be orthonormal. Then
\begin{equation}
    \sigma_{i}\left(AV^{\top}\right)= \sigma_{i}\left(A\right),
    i=1,\ldots, m.
\end{equation}
\end{lemma}

\begin{lemma}\label{lem:singular_block}
Let $A =[A_{1}, A_{2}]$. Then
\begin{equation}
    \sigma_{i}\left(A_{1}\right) \leq \sigma_{i}\left(A\right),
    i=1,\ldots, m.
\end{equation}
\end{lemma}

\begin{lemma}\label{lem:singulardot}
Let $A, B \in \mathbb{R}^{m\times n}$ and $q = \min\{m,n\}$. Then
\begin{eqnarray}
    \sigma_{i}\left(A B^\top\right) \leq \sigma_{i}(A)\sigma_1(B), i=1,\ldots,
    q.\nonumber
\end{eqnarray}
\end{lemma}

We next consider the tsm of a collection with two documents. The
following theorem comes as a generalization of a similar result of
Zha and Simon \cite[Theorem 3.3]{SimonZha.99}.
\begin{theorem}\label{*theorem*:generzhasimon}
Let $A \in \Rmn{m}{n}$ and write $A =[A_{1}, A_{2}]$. Then for any
$k_1$, $k_2$, we have
\begin{eqnarray*}
 \sigma_{i}\left([{\rm best}_{k_1}\left(A_{1}\right), {\rm best}_{k_2}\left(A_{2}\right)]\right)
\leq \sigma_{i}\left([A_{1}, A_{2}]\right).
\end{eqnarray*}
\end{theorem}

\begin{proof}
In the sequel, we remind that the  SVD of a matrix $A$ can be
written as
\begin{eqnarray}
    A = [P_k, P_k^\bot]{\rm diag}(\Sigma_k, \Sigma_k^\bot)([Q_k, Q_k^\bot])^\top, \nonumber
\end{eqnarray}
where $P_k, \Sigma_k, Q_k$ consists of the  $k$ leading left and
write singular triplets and $P_k^\bot, \Sigma_k^\bot, Q_k^\bot$
the remaining ones. Clearly, $P_k^\top P_k^\bot$ and $Q_k^\top
Q_k^\bot$ are 0 matrices. Then, for $i = 1, \ldots, m$,
$\sigma_i([A_1, A_2]) = $
\begin{eqnarray*}
\begin{array}{lcl}
    &=& \sigma_i\left( [ [P_{k_1},P_{k_1}^\bot]{\rm diag}(\Sigma_{k_1},\  \Sigma_{k_1}^\bot), A_2]\right)\\
    &=&
    \sigma_i\left([P_{k_1}\Sigma_{k_1},A_2,P_{k_1}^\bot\Sigma_{k_1}^\bot]\right)\\
    &=&
    \sigma_i\left([P_{k_1}\Sigma_{k_1}Q_{k_1}^\bot,A_2,P_{k_1}^\bot\Sigma_{k_1}^\bot]\right)({\rm by\
    Lemma\
    \ref{lem:singular_orthonormal}})\\
    &=&
    \sigma_i\left([{\rm best}_{k_1}(A_1), [P_{k_2},P_{k_2}^\bot]{\rm diag}(\Sigma_{k_2},\Sigma_{k_2}^\bot) ,P_{k_1}^\bot\Sigma_{k_1}^\bot]\right)\\
    &=&
    \sigma_i\left([{\rm best}_{k_1}(A_1),P_{k_2}\Sigma_{k_2}, P_{k_2}^\bot\Sigma_{k_2}^\bot ,P_{k_1}^\bot\Sigma_{k_1}^\bot]\right)\\
    &=&
    \sigma_i\left([{\rm best}_{k_1}(A_1),P_{k_2}\Sigma_{k_2}Q_{k_2}^\bot, P_{k_2}^\bot\Sigma_{k_2}^\bot
    ,P_{k_1}^\bot\Sigma_{k_1}^\bot]\right)\\
    &=&
    \sigma_i\left([{\rm best}_{k_1}(A_1),{\rm best}_{k_2}(A_2), P_{k_2}^\bot\Sigma_{k_2}^\bot
    ,P_{k_1}^\bot\Sigma_{k_1}^\bot]\right).
\end{array}
\end{eqnarray*}
Noticing that $[{\rm best}_{k_1}(A_1),\ {\rm best}_{k_2}(A_2)]$ is
a submatrix of\   $[{\rm best}_{k_1}(A_1),{\rm best}_{k_2}(A_2),
P_{k_2}^\bot\Sigma_{k_2}^\bot, P_{k_1}^\bot\Sigma_{k_1}^\bot]$ we
obtain the result by invoking Lemma \ref{lem:singular_block}.
\end{proof}

Using Theorem \ref{*theorem*:generzhasimon}, we can prove the
following result for the approximate tdm constructed by our
method.
\begin{theorem}\label{*theorem*:uppperbound}
Let $A \in \mathbb{R}^{m\times n}$ and write $A =[A_{1}, A_{2}]$
where $A_{1} \in \mathbb{R}^{m\times n_1}$ and $A_{2} \in
\mathbb{R}^{m\times n_2}$. Then for any $k_1$, $k_2$, we have
\begin{eqnarray}
\ \ \sigma_{i}\left([{\rm
best}_{k_1}\left(A_{1}\right)e^{(n_1)},{\rm
best}_{k_2}\left(A_{2}\right)e^{(n_2)}]\right) \\ \leq \ \ \ \ \ \
\ \sigma_{i}\left([A_{1}e^{(n_1)}, A_{2}e^{(n_2)}]\right)\nonumber
\end{eqnarray}
\end{theorem}
\begin{proof}
Let $\widehat{A} = [{\rm best}_{k_1}(A_1), {\rm
best}_{k_2}(A_2)]$. Then we have:
\begin{eqnarray}
[A_1 e^{(n_1)}, A_2 e^{(n_2)}] &=& [A_1,
A_2]\left[\begin{array}{cc}e^{(n_1)}&0\\0&e^{(n_2)}\\\end{array}\right]\nonumber\\
&=& A
\left[\begin{array}{cc}e^{(n_1)}&0\\0&e^{(n_2)}\\\end{array}\right]\nonumber
\end{eqnarray}
and
\begin{eqnarray}
[{\rm best}_{k_1}(A_1) e^{(n_1)}, {\rm best}_{k_2}(A_2) e^{(n_2)}]
&=&
\widehat{A}\left[\begin{array}{cc}e^{(n_1)}&0\\0&e^{(n_2)}\\\end{array}\right]\nonumber
\end{eqnarray}
We now define $$B^\top =
\left[\begin{array}{cc}e^{(n_1)}&0\\0&e^{(n_2)}\\\end{array}\right]$$
and by invoking Lemma \ref{lem:singulardot}, we have:

\begin{eqnarray}
\sigma_i([A_1 e^{(n_1)}, A_2 e^{(n_2)}]) \leq \sigma_i(A)
\sigma_1(B)\nonumber\\
\sigma_i([{\rm best}_{k_1}(A_1) e^{(n_1)}, {\rm best}_{k_2}(A_2)
e^{(n_2)}]) \leq \sigma_i(\widehat{A}) \sigma_1(B)\nonumber
\end{eqnarray}
As $\sigma_i(\widehat{A}) \leq \sigma_i(A)$ (by Theorem
\ref{*theorem*:generzhasimon}) the result follows.
\end{proof}
Theorem \ref{*theorem*:uppperbound} is readily generalized  to
provide bounds for the singular values of matrices $A = [A_1, A_2,
\dots A_k]$ that correspond to collection's tem's with $k$
documents.

 Note that since every term of the pseudo-tdm vector that corresponds
 to the $j$-th document is the result of a local operation  on the document's tsm,
 namely $a_j = {\rm best}_{k'} (S) e^{(t)}$,
the construction of each element of $a_j$ can be interpreted as
the result of a weight factor based on local information applied
on the corresponding term of $S e^{(t)}$.

%
%
%
\begin{center}
 \begin{table}[tbhp]
\begin{center}
 \begin{tabular}{|c||c|c|c|c|c|}
    \hline
MED &  VSM  &  New  & LSI & LSI & New w.\\
\#&  &   & $k=20$  & $k=100$ & LSI(20) \\
    \hline
1 & $0.0313$ & 0.0314  &0.0284  &0.0285  &0.0279   \\
2& $0.0754$ & 0.0728  &0.0697 & 0.0815 & 0.0633   \\
3& $0.0544$ & 0.052   &0.0595 & 0.0503 & 0.0563\\
4& $0.0793$ & 0.0797  &0.0866 & 0.0802 & 0.0919\\
5& $0.0809$ & 0.0795  &0.0888 & 0.0842 & 0.094   \\
6& $0.088$ & 0.0883  &0.0949 & 0.0898 & 0.1027   \\
7& $0.0866 $ & 0.0904 & 0.0962 & 0.0897 & 0.092   \\
8& $0.0862$ & 0.0928  &0.0938 & 0.0867 & 0.1116   \\
9& $0.1031$ & 0.1041  &0.1057 & 0.105  & 0.0992   \\
10& $0.1048$ & 0.1077 & 0.1015 & 0.1035 & 0.1016   \\
    \hline
\end{tabular}
\end{center}
 \caption{Mean precision for MEDLINE datasets. The pseudo-tdm vectors were computed with $k'=5$ singular triplets.}
 \label{localweights1}
 \end{table}
\end{center}

\section{Experimental results \label{sec:7}}
All experiments were conducted using MATLAB 6.5 running using
Windows XP on a 2.4 GHz Pentium IV PC with 512 MB of RAM. In all
cases we compute the necessary singular triplets by means of the
MATLAB {\tt svds} function that is based on implicitly restarted
Arnoldi \cite{ARPACK.98}. Our focus was query evaluation using
Equation \ref{*eq*:cosines} on the pseudo-tdm constructed via the
Algorithm of Table \ref{tb:Algo}. We tested the new method using
$k' = 1$ and $5$ tsm singular triplets to build the pseudo-tdm. We
also used the new method in combination with LSI, that is applying
LSI on the pseudo-tdm, to get an appreciation of the  overall
performance. These results were compared with simple VSM (term
frequency local weighting) and LSI using the approximated tdm of
rank $k=20, 100$. All experiments were conducted using Text to
Matrix Generator ({\sc tmg}), a recent MATLAB toolbox
\cite{TMG-inbook.06}. To this effect, we also enhanced TMG's
functionality to permit the creation of tsm's and pseudo-tdm's.\\
Tables ~\ref{localweights1} and ~\ref{localweights2} tabulate the
mean precision of VSM, LSI (based on 20 and 100 singular triplets)
and the new method for the different MEDLINE datasets. They also
illustrate the performance of the method when it is combined with
LSI
 (column ``New w.
 LSI'' of Tables ~\ref{localweights1}, ~\ref{localweights2}).

\begin{center}
 \begin{table}[tbhp]
\begin{center}
 \begin{tabular}{|c||c|c|c|c|c|}
    \hline
MED &  VSM  &  New  & LSI & LSI & New w.\\
\#&  &   & $k=20$  & $k=100$ & LSI(20) \\
    \hline
 1 & 0.0313 & 0.0325  &0.0284  &0.0285  &0.0283
\\
 2& 0.0754  &0.0657  &0.0697  &0.0815  &0.0576
\\
 3& 0.0544  &0.0527  &0.0595  &0.0503  &0.0534
\\
 4& 0.0793  &0.077   &0.0866  &0.0802  &0.0779
\\
 5& 0.0809  &0.0792  &0.0888  &0.0842  &0.0809
\\
 6& 0.088   &0.0816  &0.0949  &0.0898  &0.0903
\\
 7& 0.0866  &0.0903  &0.0962  &0.0897  &0.0948
\\
 8& 0.0862  &0.0867  &0.0938  &0.0867  &0.103
\\
 9& 0.1031  &0.1008  &0.1057  &0.105   &0.0949
\\
 10& 0.1048  &0.1006  &0.1015  &0.1035 & 0.1005
   \\
    \hline
\end{tabular}
\end{center}
 \caption{Mean precision for
  MEDLINE datasets. The pseudo-tdm vectors were computed with $k'=1$ singular triplets.}
 \label{localweights2}
 \end{table}
\end{center}
Tables ~\ref{numberqueries5}, ~\ref{numberqueries1} present the
number of queries that each method answers with greater precision,
compared to the precision of the other method's answers. The new
method appears to offer significant  improvements over the
performance of VSM, while in many cases the new methods performs
better than LSI.
\begin{table}[tbh]
\begin{center}
 \begin{tabular}{|c||c|c|c|c|c|}
    \hline
    MED &  VSM  &  New  & LSI & LSI \\
\#&  &   & $k=20$  & $k=100$  \\
    \hline
 1& 2(7\%)  & 9(30\%) &  9(30\%) & 7(23\%)
\\
 2& 3(10\%)  & 8(27\%)  &9(30\%) & 9(30\%)
\\
 3& 4(13\%)   &3(10\%) &14(47\%) & 8(27\%)
\\
 4& 4(13\%)  &7(23\%) & 15(50\%) & 4(13\%)
\\
 5& 4(13\%) &6(20\%) & 13(43\%) & 5(17\%)
\\
 6& 3(10\%) &7(23\%) & 11(37\%) & 7(23\%)
\\
 7& 4(13\%)   &9(30\%) & 13(43\%) & 2(7\%)
\\
 8& 3(10\%) &5(17\%) & 14(47\%) &7(23\%)
\\
 9& 2(67\%) &7(23\%) &9(30\%) & 10(33\%)
\\
 10& 6(20\%) & 9(30\%) & 13(43\%) & 1(3\%)
   \\
    \hline
\end{tabular}
\end{center}
 \caption{Number of queries that each method answers with greater precision for MEDLINE datasets.
 The pseudo-tdm vectors were computed with $k'=5$ singular triplets. }
 \label{numberqueries5}
 \end{table}

\begin{figure}[htbp]
\begin{center}
\includegraphics[width=250pt,height=130pt,angle=0]{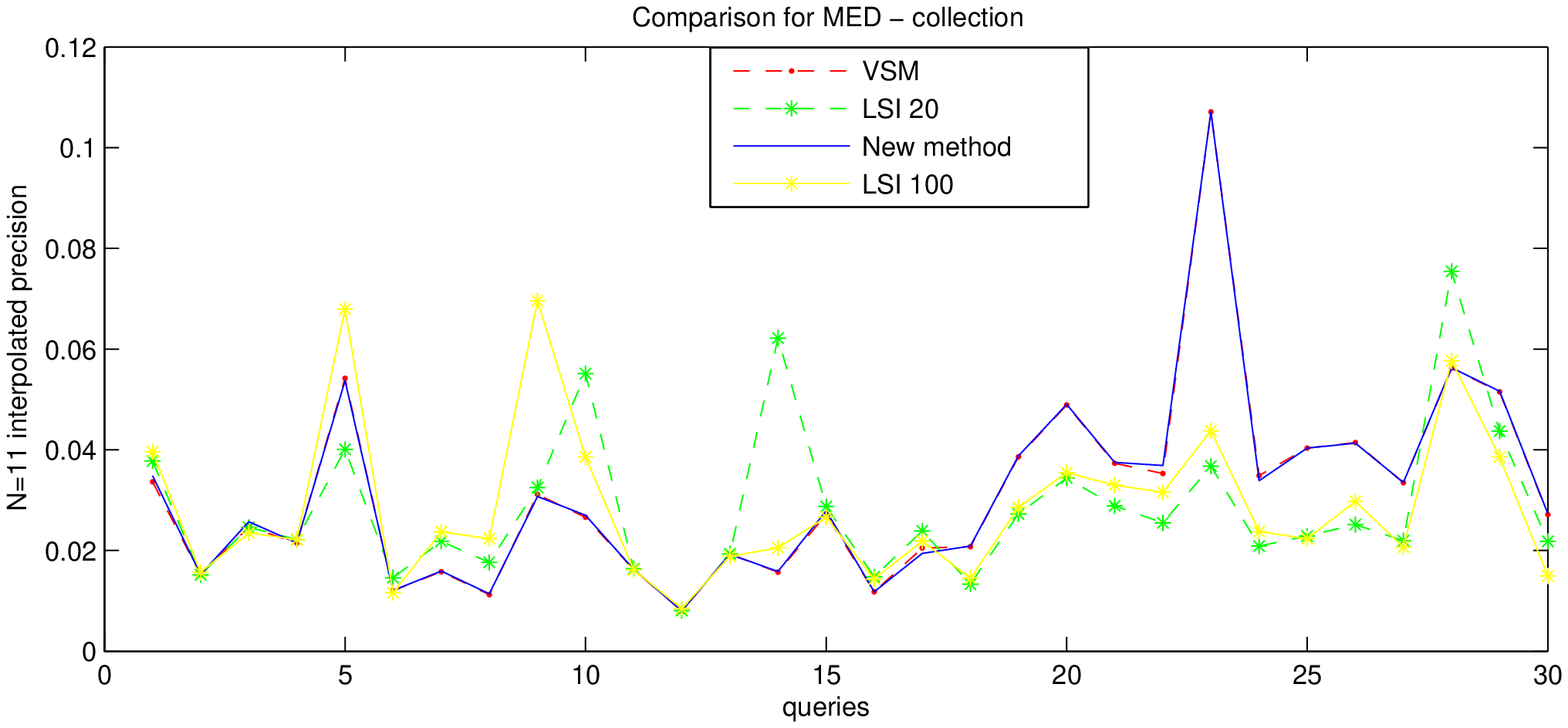}
\includegraphics[width=250pt,height=130pt,angle=0]{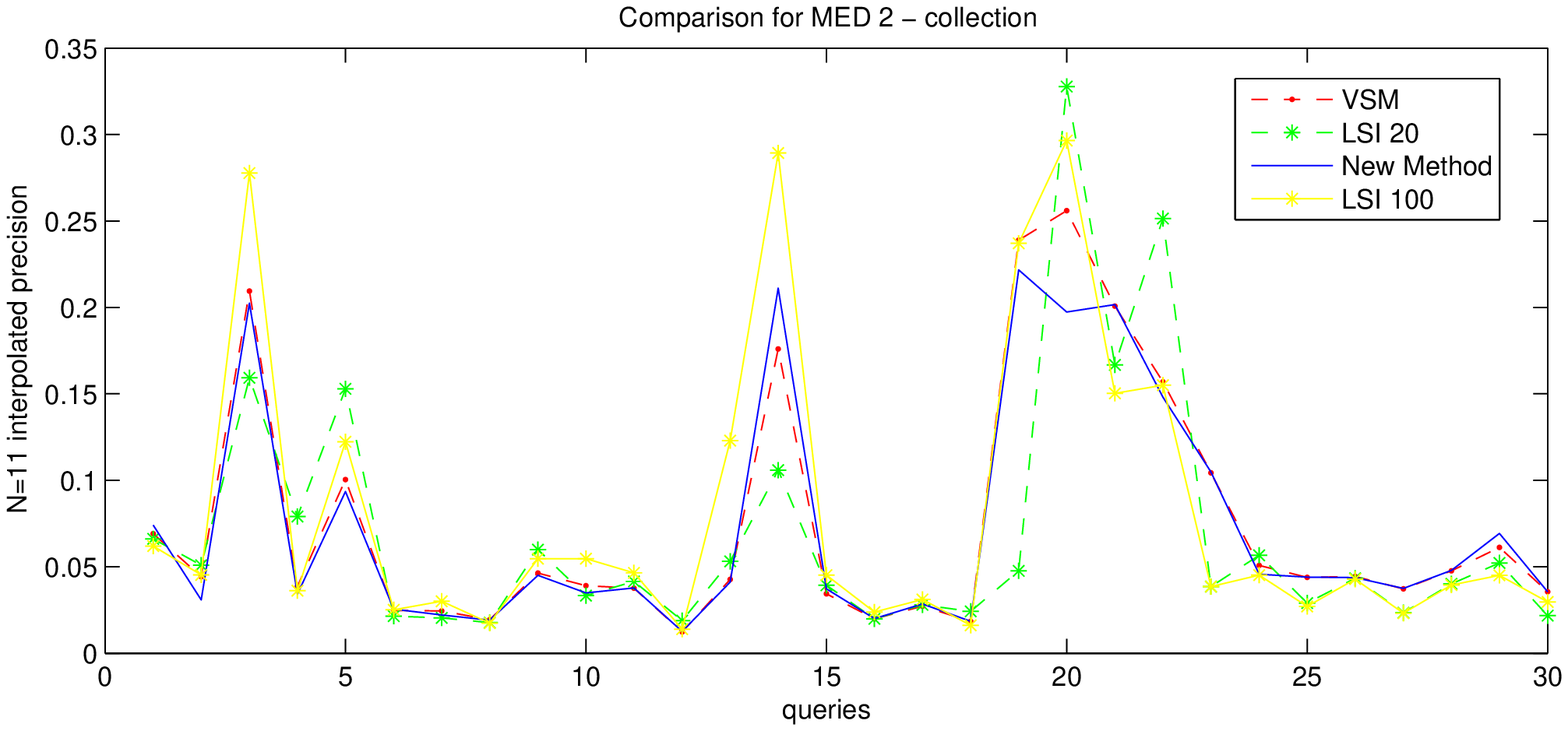}
\includegraphics[width=250pt,height=130pt,angle=0]{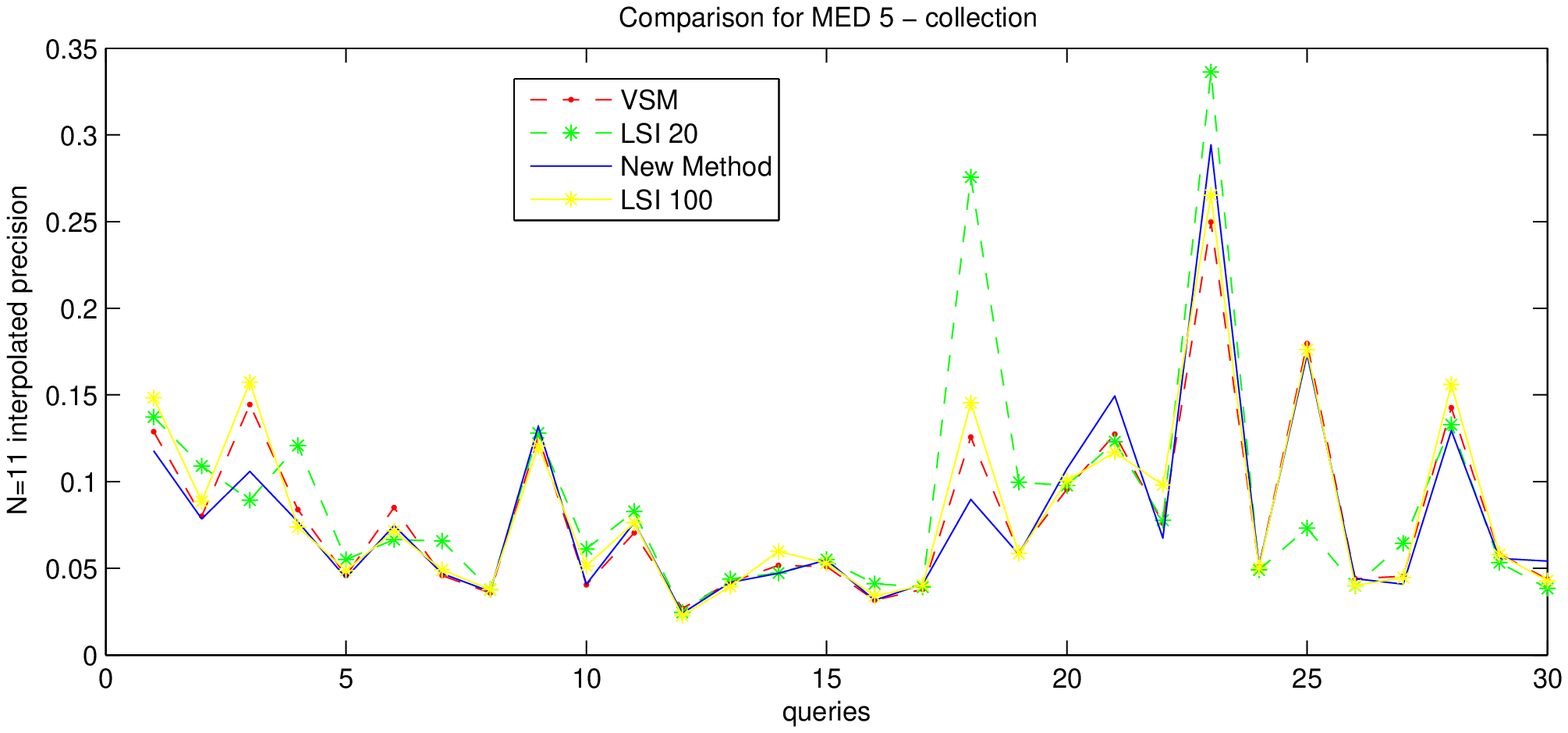}
\includegraphics[width=250pt,height=130pt,angle=0]{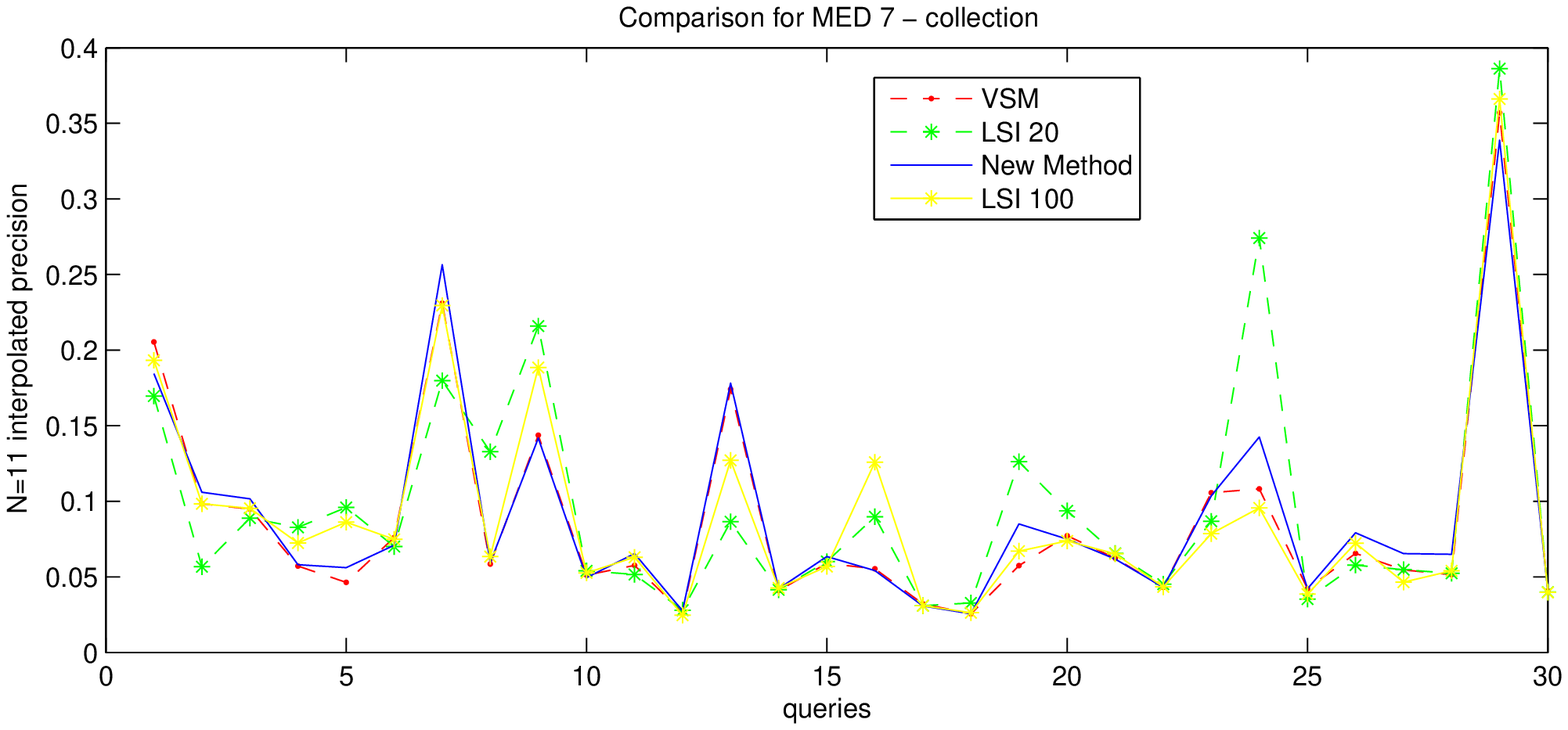}
\includegraphics[width=250pt,height=130pt,angle=0]{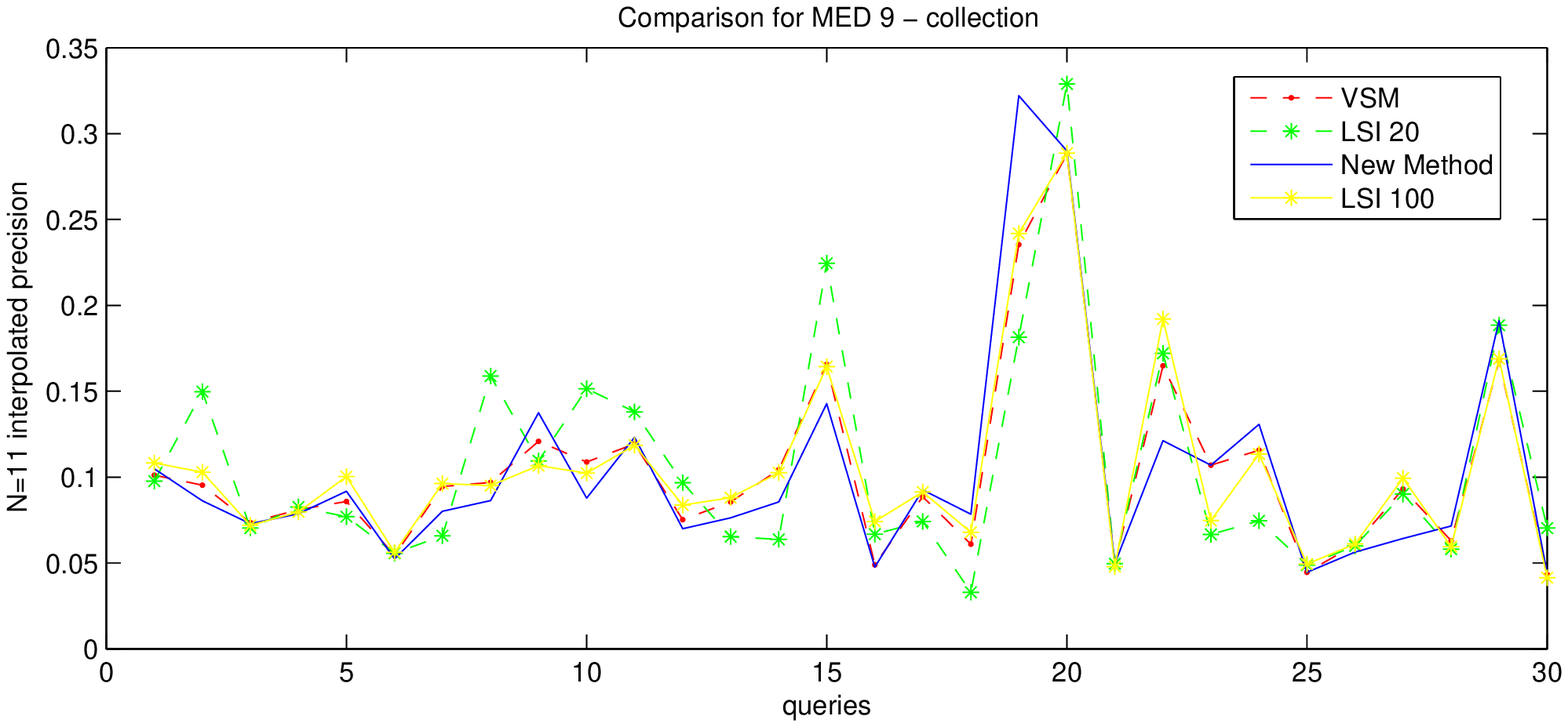}
\caption{ {\small  Results for the new method using 5 singular
triplets to approximate the tsm, classic VSM and LSI with $20$ and
 $100$ singular triplets, applied on MEDLINE datasets.} }
\label{FS2}
\end{center}
\end{figure}

\begin{table}[tbh]
\begin{center}
 \begin{tabular}{|c||c|c|c|c|c|}
    \hline
    MED &  VSM  &  New  & LSI & LSI \\
\#&  &   & $k=20$  & $k=100$  \\
    \hline
 1 & 1(3\%)  & 15(50\%) & 7(23\%)   &  6(20\%)
\\
 2& 4(7\%)  & 8(27\%)  & 9(30\%)  & 9(30\%)
\\
 3& 3(10\%)  & 7(23\%)  & 14(47\%) & 6(20\%)
\\
 4& 4(13\%) &  6(20\%) & 14(47\%) & 5(17\%)
\\
 5& 4(13\%)  & 6(20\%)  & 14(47\%) & 5(17\%)
\\
 6& 4(13\%)  & 9(30\%)  & 8(27\%)  & 7(23\%)
\\
 7& 5(17\%) & 8(27\%)  & 11(37\%) & 4(13\%)
\\
 8& 3(10\%)  & 7(23\%)   &13(43\%) & 5(17\%)
\\
 9&1(3\%)  & 8(27\%)   &8(27\%)  & 11(37\%)
  \\
10& 7(23\%) &  8(27\%) &  12(40\%) & 3(10\%)
   \\
    \hline
\end{tabular}
\end{center}
 \caption{Number of queries that each method answers with greater precision, for MEDLINE datasets.
The pseudo-tdm vectors were computed with $k'=1$ singular
triplets. }
 \label{numberqueries1}
 \end{table}
We also plot, in Figure \ref{FS2}, the $N=11$-point interpolated
average precision for the different queries of MEDLINE datasets.
The interpolated precision is defined as:
\begin{eqnarray}
    p = \frac{1}{N}\sum_{i=0}^N \widehat{p}\left(
    \frac{i}{N-1}\right)\nonumber
\label{*eq*:npointprecision}
\end{eqnarray}
where:
\begin{eqnarray}
    \widehat{p}\left(x\right) = \max \{p_i\ \mid\  n_i \geq xr\ , i=1:r
    \}\nonumber
\label{*eq*:precisionatrecalllevel}
\end{eqnarray}

\begin{figure}[htbp]
\begin{center}
\includegraphics[scale=.40,angle=0]{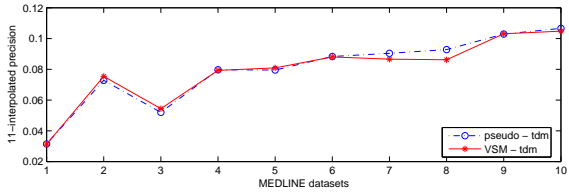}
\caption{ {\small VSM vs approximated} }
\label{VSMvsapproximatedtotal.}
\end{center}
\end{figure}
\begin{center}
\begin{figure}[htbp]
\begin{center}
\includegraphics[scale=.40,angle=0]{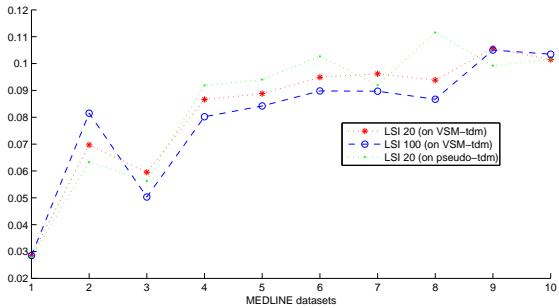}
\caption{ {\small The new method as an alternative local weighting
scheme. LSI based on the pseudo-tdm has better performance than
LSI on the VSM tdm.} } \label{LSI.}
\end{center}
\end{figure}
\end{center}
\begin{center}
\begin{figure}[htbp]
\begin{center}
\includegraphics[scale=.40,angle=0]{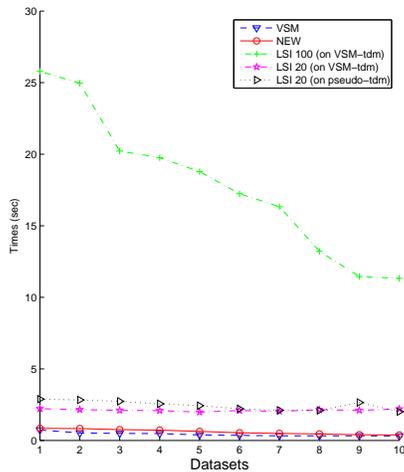}
\caption { {\small Runtimes (seconds) for methods discussed in
this paper.} } \label{fig:times.eps}
\end{center}
\end{figure}
\end{center}

Figure \ref{VSMvsapproximatedtotal.} illustrates the performance
of the new method (using
 5 singular triplets to approximate the tsms) compared to VSM. Figure ~\ref{LSI.}
 provides experimental results for the new method viewed as an
alternative weighting scheme. The new method and VSM have similar
performance on datasets MED\_1 to MED\_5. However, for MED\_6 to
MED\_10, the new method improves VSM. These results imply that the
SVD approximation of tsm's indeed captures the topic directions of
multi-topic documents and thus  improves the overall IR
performance. Furthermore, LSI's performance improves when based
upon the pseudo-tdm.\\
 Finally, in order to gain an appreciation
for the method's cost, we present in Figure \ref{fig:times.eps}
 the runtimes for performing a query on tdm's that correspond to
classical VSM, LSI with values of $k=20$ and 100, the new method,
and finally the combination of LSI with the new method. In all
experiments, times include the  cost of performing the necessary
partial SVD's. Results indicate  the new method has runtimes
similar to  VSM.

\section{Conclusions \label{sec:8}}

Our theoretical and experimental results suggest that,  at the
sentence level,  matrix space models that have greater awareness
of each document's local structure and are able to capture
additional semantic  information for each document, can be
successfully used to improve existing IR techniques. Our results
provide significant evidence that  further justify proposals such
as those in \cite{HammoudaKamel.04,Roelleke-etal2004} towards the
use of matrix based models and provide additional tools for IR in
such frameworks.
%
%
 We are
currently studying the effects of enabling additional levels of
analysis (not only based on sentences) and adding overall greater
flexibility in the algorithm, as well as the utilization of
multilinear algebra techniques (cf. \cite{LZYCLBC05}) and the use
of parallel processing.

\subsection*{Acknowledgments.} We thank the referees for their suggestions.

\bibliography{EG,IR,IRbib}

\end{document}